# Vortex Core and POD Analysis on Hairpin Vortex Formation in Flow Transition

Sita Charkrit and Chaoqun Liu

*Department of Mathematics, University of Texas at Arlington, Arlington, Texas 76019, USA*

In this study, the new concept of vortex core line based on Liutex definition is applied to demonstrate that vortex ring is not part of the Λ-vortex and the generation of ring-like vortex is formed separately from the Λ-vortex. The proper orthogonal decomposition (POD) is also applied to analyze the Kelvin–Helmholtz (K-H) instability happening in hairpin ring areas of flow transition on the flat plate to understand the ring-like vortex formation. The new identification method named modified Omega-Liutex method is efficiently used to visualize and observe the shapes of vortex structures in 3-D. The streamwise vortex structure characteristics can be found in POD Mode1 as the mean flow. The other POD modes are in spanwise structures and have the fluctuation motions, which are induced by K-H instability. Moreover, the result shows that fluctuated POD modes are in pairs and share the same characteristics such as amplitudes, mode shapes and time evolutions. The vortex core and POD results confirm that the Λ-vortex is not self-deformed to a hairpin vortex, but it is formed by the K-H instability during the formation of Λ-vortex and hairpin vortex in boundary layer flow transition.

## Nomenclature

$M_\infty$ = Mach number
$Re$ = Reynolds number
$x_{in}$ = distance between leading edge of flat plate and upstream boundary of computational domain
$\delta_{in}$ = inflow displacement thickness
$Lx$ = length of computational domain along x direction
$Ly$ = length of computational domain along y direction
$Lz_{in}$ = height at inflow boundary
$T_w$ = wall temperature
$T_\infty$ = free stream temperature

## 1. Introduction

Vortices are considered as a building block of turbulent flows. Vortices can be characterized by the structure shapes, which can be observed in experiments or computations. It is well-known that spanwise vortex (parallel to the y-axis), Λ-shaped vortex, and hairpin vortex commonly appear in flow transition in both experiment and DNS simulation. The hairpin vortex typically consists of three parts: 1) two counter-rotating legs; 2) a ring-like vortex known as the vortex head, which is always in Ω-shape; 3) necks that connect the head and two legs. The formation of vortex is an important area to study the physics of vortex and turbulence development. To study how one type of vortex becomes another type especially in flow transition is one way to get better understanding about turbulence. In flow transition, a spanwise vortex is formed at the very beginning, then the Λ-vortex appears. It was found in the studies [1, 2] that the mechanism of a spanwise vortex becomes a Λ-vortex is due to the vortex filaments redirection and reorganization. The two roots of a Λ-vortex known as two legs are two counter rotating centers containing opposite directions of rotation. After the Λ-shaped vortex is formed, the hairpin vortex appears. Then, multiple vortex rings are formed one by one.

For the mechanism of the Λ-vortex becoming the hairpin vortex, many researchers believe that the formation of vortex ring is caused by vorticity self-deformation. Hama [3] and Hama and Nutant [4] studied the formation and development of Λ-shaped vortices by applying means of flow visualization in water. Knapp and Roache [5] also concluded that the self-deformation of Λ-vortex to hairpin vortex has been observed by means of smoke visualizations in air. Moin et al. [6] gave an explanation about the mechanism for the generation of ring-like vortices in turbulent shear flows by using Biot-Savart law for 2-D and DNS for 3-D. The results in [6] demonstrated that "perturbation of layer of vorticity leads to its roll-up into filaments of concentrated vorticity" or, equivalently, the Λ-vortex becomes the hairpin vortex by self-deformation as shown in figure 1. Although the calculation of vortex layer in the study [6] could be correct, the concept of vortex definition could be incorrect as the vorticity lines were used to represent vortex tubes. This would lead to the misunderstanding about the hairpin vortex formation through the self-deformation of the Λ-vortex.

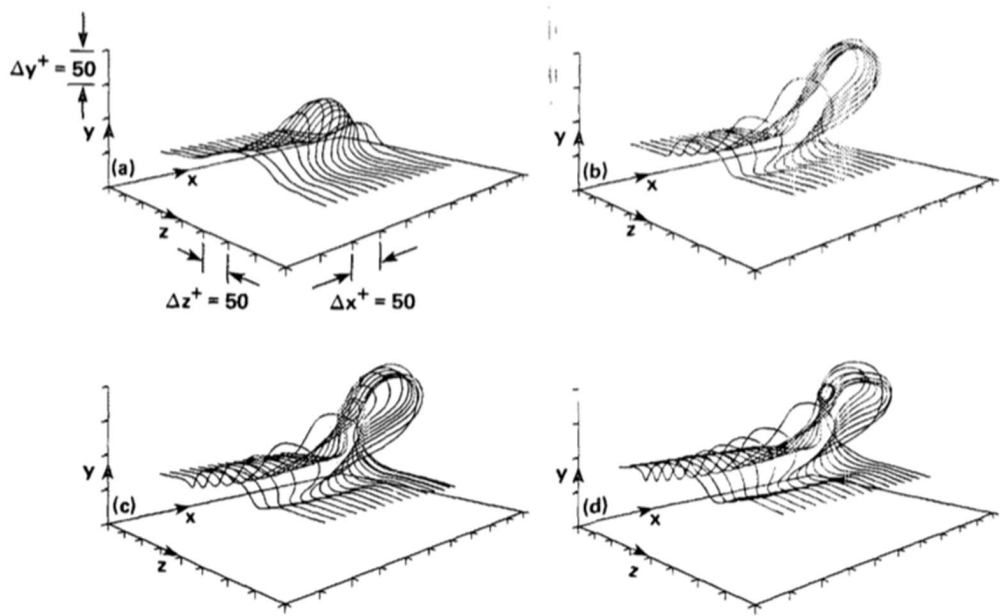

**Figure 1.** Sketch of vortex layer at different time from (a) to (d) [6]

Since Helmholtz [7] introduced the concepts of vorticity filament and tube in 1858, many people have considered to use a vorticity tube to represent a vortex. The vorticity is mathematically defined by velocity curl, $\boldsymbol{\omega} = \boldsymbol{\nabla} \times \mathbf{V}$, and the vortex filament known as the vorticity line, which is defined as the line everywhere tangent to the local vorticity vector $\boldsymbol{\omega}$, has been mistakenly considered as the infinitesimal vortex tube. However, this concept could cause many misunderstandings and may be one of the major obstacles in turbulence research. Vorticity tubes cannot represent vortices in general, especially in viscous flows. Robinson [8] demonstrated that the actual vortices can be weak even though the vorticity is strong. Since then, many vortex identification methods have been introduced to capture the vortex structures. One of the popular and well-known vortex identification criteria named as $\lambda_2$ [9] has been widely applied in computational fluid dynamic community. However, the physical meaning of $\lambda_2$ is not very clear and the iso-surface of $\lambda_2$ could present some fake vortex structures depending on the selection of the threshold [1, 2]. Therefore, the use of $\lambda_2$-criterion may not be appropriate to visualize vortex structures.

In 2014, the combination of $\lambda_2$ and vortex filament was applied by Yan et al. [1] and Liu et al. [2] to study the mechanism of the late stages of flow transition in a boundary layer at a free stream Mach number of 0.5 using DNS simulation. The difference between vortex filament that Helmholtz defined and natural vortex was also pointed out. The vortex legs are only a concentration of more local vortex filaments. The Λ-vortex and subsequent legs of hairpin vortex are not vorticity tubes since a vorticity tube cannot be penetrated by any vorticity lines. By the method of combination of $\lambda_2$ and vorticity lines, as a result, they gave a contradiction to the previous conclusions of self-deformation of the Λ-vortex to hairpin vortex. Then, the new mechanism of ring-like vortex formations was revealed that there is no such a process that a Λ-vortex is self-deformed to a hairpin vortex as many literatures suggested. Actually, the vortex ring is not a part of the original Λ-vortex but is formed separately, and the formation of a Λ-vortex to a hairpin vortex is caused by the shear layer instability or so-called K-H instability. The mechanism of the formation of a Λ-vortex to a hairpin vortex caused by K-H instability will be revisited in Section 4.

Recently, Liu et al [10, 11] and Gao and Liu [12] introduced a new idea of "Liutex" to represent a rotation part of fluid flow. As the previous discussion, vortices cannot be represented by the vorticity in general. Actually, vorticity should be decomposed into two parts, a rotational part and a non-rotational part. In other words, vorticity consists of Liutex ($\boldsymbol{R}$) and the antisymmetric shear ($\boldsymbol{S}$) described by the mathematical expression $\boldsymbol{\omega} = \nabla \times \boldsymbol{V} = \boldsymbol{R} + \boldsymbol{S}.$ Moreover, the velocity gradient tensor can be decomposed to a rotational part ($\boldsymbol{R}$) and a non-rotational part ($\boldsymbol{NR}$), i.e., $\nabla \boldsymbol{V} = \boldsymbol{R} + \boldsymbol{NR}$. Liutex vector is the rotational part of vorticity without shear contamination and defined by $\boldsymbol{R} = R\boldsymbol{r},$ where $\boldsymbol{r}$ is the eigenvector of the velocity gradient tensor $\nabla \boldsymbol{V}$ and $R$ is the rotation strength (the rigid-body angular speed). Wang et al. [13] recently proposed an explicit formula of the magnitude $R$ as follows.

$$R = \boldsymbol{\omega} \cdot \boldsymbol{r} - \sqrt{(\boldsymbol{\omega} \cdot \boldsymbol{r})^2 - 4\lambda_{ci}^2},$$

where $\boldsymbol{\omega}$ is a local vorticity vector and $\lambda_{ci}$ is an imaginary part of the complex eigenvalue of $\nabla \boldsymbol{V}.$ As the above description, Liutex is able to provide the rotation axis, the rigid-body angular speed and the tensor form. In addition, Liutex is local, accurate, unique, systematical and Galilean invariance. Since vortices are recognized as the rotational motion of fluids, it is more reasonable to represent vortex by Liutex than using vorticity or any other vortex identification criteria.

After the Liutex concept was proposed, the new vortex identification method named as modified Omega-Liutex was developed by Liu and Liu [14] to improve the existing vortex identification methods such as $Q$-criterion and $\lambda_2$-criterion. This method is a dimensionless relative quantity from 0 to 1. According to the study [14], the new modified method has many advantages as follows: 1) It can distinguish the vortex from high shear layer boundary layers; 2) It is robust and can be set as 0.52 to approximate the iso-surface of vortex structure; 3) It can capture both weak and strong vortices simultaneously. Moreover, Gao et al. [15] has recently introduced the identification of vortex core lines based on Liutex definition. It can represent the vortex rotation axis line, which requires that the Liutex magnitude gradient vector is aligned with the Liutex vector. This vortex core line can be obtained as a unique line identifying the vortex core structure instead of using the iso-surface.

For more accurate and more reasonable methods of vortex identification, several tools based on Liutex: Liutex magnitude, modified Omega-Liutex method and vortex core line, are efficiently used to analyze vortex structures in this paper. Liutex magnitude and modified Omega-Liutex method will be applied in this study to capture vortex structures in 3-D with iso-surface, as well as the vortex core lines based on Liutex will be used to analyze the formation of ring-like vortex.

Furthermore, as the results in the studies [1, 2] have shown that a Λ-vortex is not self-deformed to a hairpin vortex (also shown by vortex core lines in our study) but is formed by K-H instability, the proper orthogonal decomposition (POD) is also used to analyze the ring-like vortex formation in flow transition. It will be applied to support the studies [1, 2] that the formation process of a Λ-vortex to a hairpin vortex is caused by K-H instability. In addition, the modified Omega-Liutex method is applied to capture the vortices of POD modes in our study in order to effectively visualize vortices and obtain more structures in both weak and strong vortices simultaneously.

POD has been widely applied to extract the coherent structure of fluid flows to understand the complexity of turbulent flow. The original POD was introduced by Lumley [16] in 1967 to investigate the turbulent flow. The other version, snapshot POD, was introduced by Sirovich [17] in 1987 to optimize the computation. In POD method, the flow structure is decomposed into orthogonal modes ranking by their kinetic energy content and it can decouple the spatial and temporal in different applications. For examples, a mixing layer downstream on a thick splitter plate obtain from DNS was examined by POD [18]. POD was also applied to turbulent pipe flow [19-22]. The vortex structure in a MVG (micro vortex generator) wake was analyzed by POD in the study [23]. POD has been applied to flow transition. For instance, a transitional boundary layer with and without control was studied by POD [24]. The vortex structure in late-stage transition was analyzed by using the POD to observe mode-shaped structures [25]. The study [26] also applied POD to investigate asymmetric structures in late flow transition. However, no one has studied the mechanism of the vortex ring formation in flow transition by use of the POD method yet.

Therefore, this paper is proposed to point out two main results by two different strategies: 1) The vortex core line based on Liutex is applied to demonstrate that the ring-like vortex is not part of the Λ-vortex and the Λ-vortex is not self-deformed to hairpin vortex; 2) The POD is used to support the hypothesis that the hairpin vortex is formed by K-H instability. In addition, this paper is organized as follows: In Section 2, the case set up and code validation is described. In Section 3, the vortex core line and the modified Omega-Liutex method are introduced as the new vortex identification methods used in this paper. In Section 4, a ring-like vortex formation is presented by the use of vortex core lines and the result of K-H instability is revisited. In Section 5, a review of POD method is explained and our results by POD are reported. Finally, the conclusion of the formation of hairpin vortex and POD analysis about K-H instability is given in the last section.

## 2. Case set up and code validation

The case and code validation has been introduced in our previous researches [1, 2, 27]. The DNS code (DNSUTA) was validated by researchers from UTA and NASA Langley. The results were compared to experiments and other's DNS results [28, 29] and the consistence shows that the code is correct and accurate.

Figures 2 and 3 present the domain of computation. The grid level is 1920 x 128 x 241, representing the number of grids in streamwise ($x$), spanwise ($y$), and wall normal ($z$) directions. The grid is stretched in the normal direction and uniform in the streamwise and spanwise directions. The length of the first grid interval in the normal direction at the entrance is found to be 0.43 in wall units ($Z+ = 0.43$). The flow parameters, including Mach number, Reynolds number, etc. are listed in Table 1. For more detail about case setup and code validation, see [1, 2].

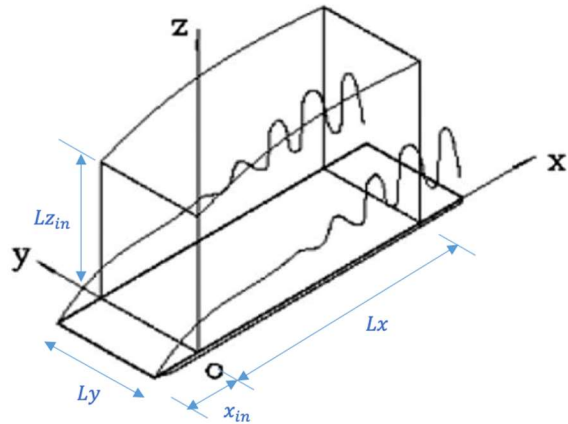

**Figure 2.** Computation domain.

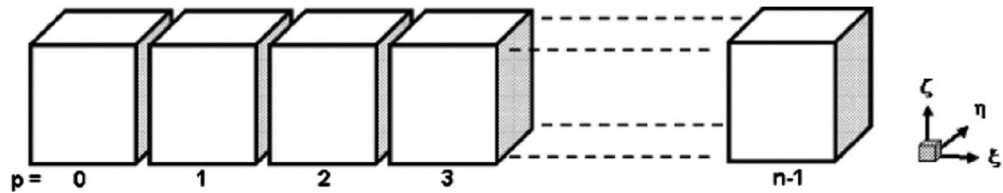

**Figure 3.** Domain decomposition along the streamwise direction in the computational space.

| $M_\infty$ | $Re$ | $x_{in}$ | $Lx$ | $Ly$ | $Lz_{in}$ | $T_w$ | $T_\infty$ |
|---|---|---|---|---|---|---|---|
| 0.5 | 1000 | $300.79\delta_{in}$ | $798.03\delta_{in}$ | $22\delta_{in}$ | $40\delta_{in}$ | 273.15K | 273.15K |

**Table 1.** DNS parameters

## 3. Vortex Core Line and Modified Omega-Liutex Method

In this section, the new concept of vortex core line and the modified Omega-Liutex method ($\widetilde{\Omega}_L$ method) proposed by Gao et al [15] and Liu and Liu [14] are reviewed.

### 3.1 Vortex core identification

There have been many vortex identification methods used as convenient tools to capture vortices. The Liutex magnitude is one of vortex identification criteria. The vortex structure in early transition stage visualized by different Liutex magnitudes are shown in figure 4. The iso-surface of Liutex magnitude can be visualized by adjusting thresholds. For instance, Liutex magnitude is set by $R = 0.1$, $R = 0.05$ and $R = 0.02$ in figures 4(a), 4(b) and 4(c), respectively. A smaller Liutex magnitude threshold can represent more vortex structures. The variety of thresholds can greatly change the appearance of the results. However, no one knows about the best specific threshold to represent vortices. This problem may lead the wrong understanding about the shapes and formations of vortex structures. Note that figures 4 (a), 4(b), 4(c) were drawn by the same DNS data but different thresholds which give totally different structures. All currently popular vortex identification methods could use the same data to give different vortex structures.

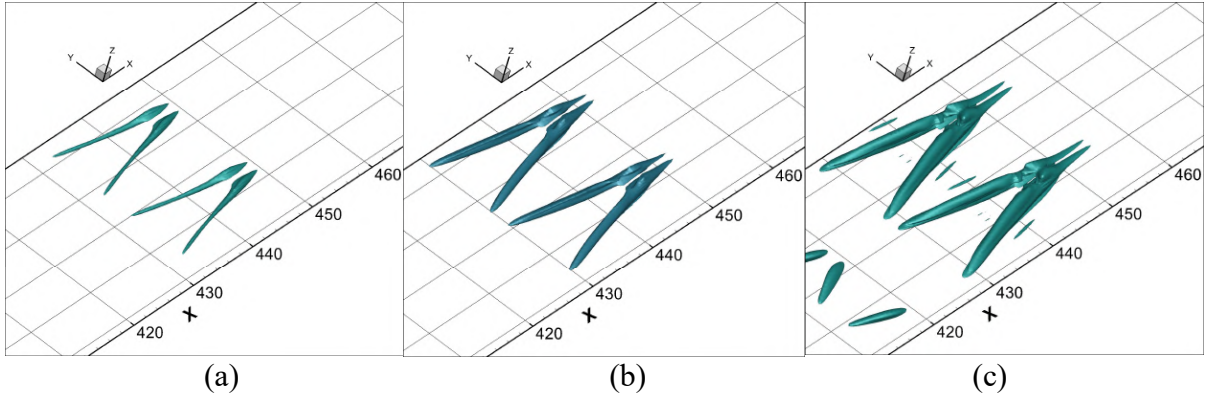

(a)          (b)          (c)

**Figure 4.** The Λ-vortex at $t = 6.0T$, where $T$ is the period of T-S wave, visualized by Liutex magnitude $R$.
(a) Iso-surface of $R = 0.1$ (b) Iso-surface of $R = 0.05$ (c) Iso-surface of $R = 0.02$.

Since the vortex core based Liutex has been mathematically proposed, it can be used to represent the vortex alternatively. According to [11, 15], some important definitions are introduced as follows.

**Definition 1:** The rotation vector named as Liutex vector ($\boldsymbol{R}$) is defined as the local rotation axis and the magnitude ($R$) of Liutex vector is defined as the rotation strength (the rigid-body angular speed) as follows.

$$\boldsymbol{R} = R\boldsymbol{r},$$

where $r$ is the eigenvector of the velocity gradient tensor $\nabla V$. The explicit formula of the magnitude $R$ [13] is given by

$$R = \boldsymbol{\omega}\cdot\boldsymbol{r} - \sqrt{(\boldsymbol{\omega}\cdot\boldsymbol{r})^2 - 4\lambda_{ci}^2},$$

where $\boldsymbol{\omega}$ is a local vorticity vector and $\lambda_{ci}$ is an imaginary part of the complex eigenvalue of $\nabla V$.
**Definition 2:** The vortex core line or the vortex rotation axis is defined as a special Liutex line passing the points that satisfy the condition

$$\nabla R \times \boldsymbol{r} = 0, R > 0,$$

where $r$ represents the direction of the Liutex vector.

As the above definitions, the vortex rotation core lines of the flow field is uniquely defined without any threshold requirement. The vortex core lines can be used to represent the vortex without iso-surface. According to [15], the vortex core lines can be extracted by a three-step manual method:
Step 1: Set a slice that shows the contour of the Liutex magnitude.
Step 2: Determine some Liutex magnitude gradient lines passing through the slice. It is to be observed that a few Liutex magnitude gradient lines is enough for each individual vortex.
Step 3: Find the intersection point of the concentration line and the slice and create a Liutex line passing through the intersection point. These special Liutex lines will be illustrated as a vortex rotation axis line.

Figure 5(a) demonstrates Liutex magnitude gradient lines, which are obtained from Step 2, of vortex shown in figure 4. Then, the vortex rotation core lines, which are obtained from Step 3, are illustrated and shown in figure 5(b). Moreover, the visualization of vortex with different thresholds can be represented by the unique vortex core lines as shown in figures 6(a), 6(b) and 6(c). Keep in mind that figures 6(a), 6(b) and 6(c) give totally different vortex structure, but they have the same vortex core lines which are unique.

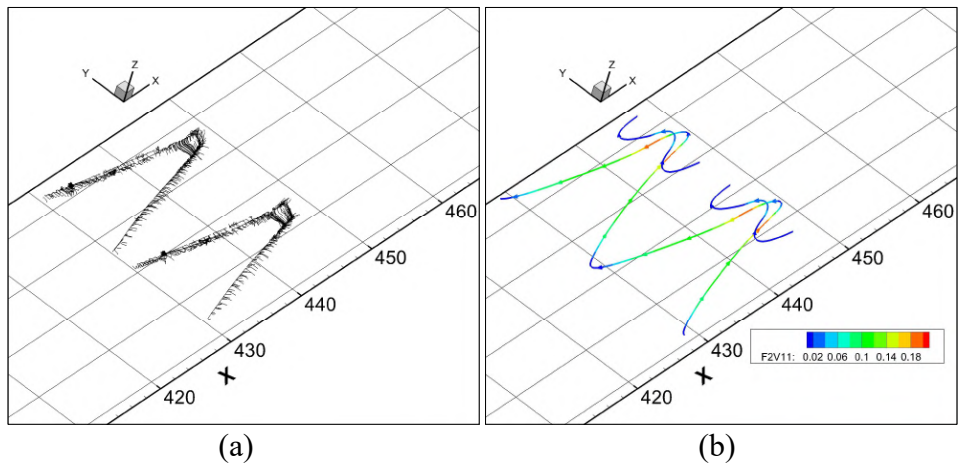

(a)          (b)

**Figure 5.** The Λ-vortex at $t = 6.0T$, where $T$ is the period of T-S wave. (a) Liutex magnitude gradient lines without iso-surface (b) Vortex core lines colored (red is strong and blue is weak) by Liutex magnitude.

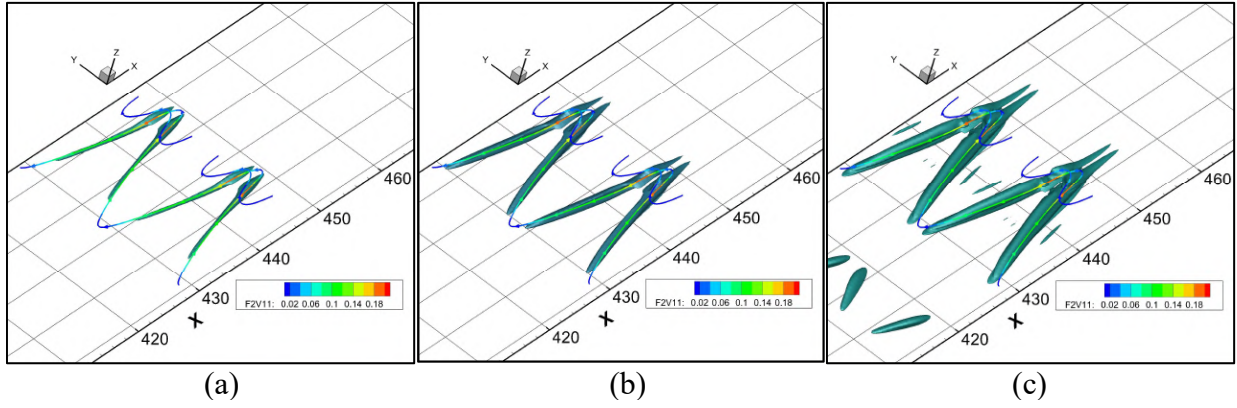

**Figure 6.** The Λ-vortex at $t = 6.0T$, where $T$ is the period of T-S wave, visualized by Liutex magnitude and vortex core line. (a) Iso-surface of $R = 0.1$ (b) Iso-surface of $R = 0.05$ (c) Iso-surface of $R = 0.02$.

## 3.2 Modified Omega-Liutex method

The new vortex identification method named modified Omega-Liutex, $\widetilde{\Omega}_L$, method has been proposed in 2019 by Liu and Liu [14] as an effective method to capture vortices. It was developed to replace the previously exiting methods such as $Q$-criterion [30] and $\lambda_2$-criterion [9], and improve the Omega method [31, 32] and Omega-Liutex method [33].

Since the Omega method has been introduced in 2016 [31, 32], many studies [31-38] have shown that it is more powerful to capture vortices than the existing vortex identification methods. After that, the Omega-Liutex method has been proposed as a tool based on Liutex definition. This method is also powerful to visualize vortices but there are some bulging phenomena on the iso-surfaces. As a result, modified Omega-Liutex method was introduced to resolve some deficiencies on the Omega-Liutex method and others.

The modified Omega-Liutex, $\widetilde{\Omega}_L$, is defined by

$$\widetilde{\Omega}_L = \frac{\beta^2}{\beta^2 + \alpha^2 + \lambda_{cr}^2 + \frac{1}{2}\lambda_r^2 + \varepsilon},$$

$$\beta = \frac{1}{2}\boldsymbol{\omega}\cdot\boldsymbol{r},$$

$$\alpha = \frac{1}{2}\sqrt{(\boldsymbol{\omega}\cdot\boldsymbol{r})^2 - 4\lambda_{ci}^2},$$

where $\boldsymbol{\omega}$ is a local vorticity vector; $\lambda_r, \lambda_{cr}$ and $\lambda_{ci}$ are a real eigenvalue, a real part of the complex eigenvalue and an imaginary part of the complex eigenvalue of $\boldsymbol{\nabla V}$, respectively. Here, $\varepsilon$ is a small positive number used to avoid division by zero and $\varepsilon = b(\beta^2 - \alpha^2)_{max}$, where $b$ is a small positive number around 0.001~0.002. In this paper, $b$ is set by $b = 0.001$ and $\widetilde{\Omega}_L = 0.52$ is applied as an empirical value according to [14].

## 4. Ring-Like Vortex Formation

It is widely recognized that the evolution of vortex formation in flow transition can be displayed as in figure 7. The illustration in figure 7 are visualized by the modified Omega-Liutex method. It can be observed that the vortex formation starts with the spanwise vortex, and then the Λ-vortex is generated. After the Λ-vortex is formed, the ring-like vortex will appear. Then, the multiple hairpin vortices are formed. In the process that the Λ-vortex becomes ring-like vortex, which can be seen in the area between $x = 450$ and $x = 470$ in the streamwise direction as shown in figure 7, it is widely believed that the Λ-vortex self-deforms to the hairpin vortex. There have been many studies [3-6] with experiments and DNS which concluded that the process of the Λ-vortex becomes hairpin vortex is through the self-deformation.

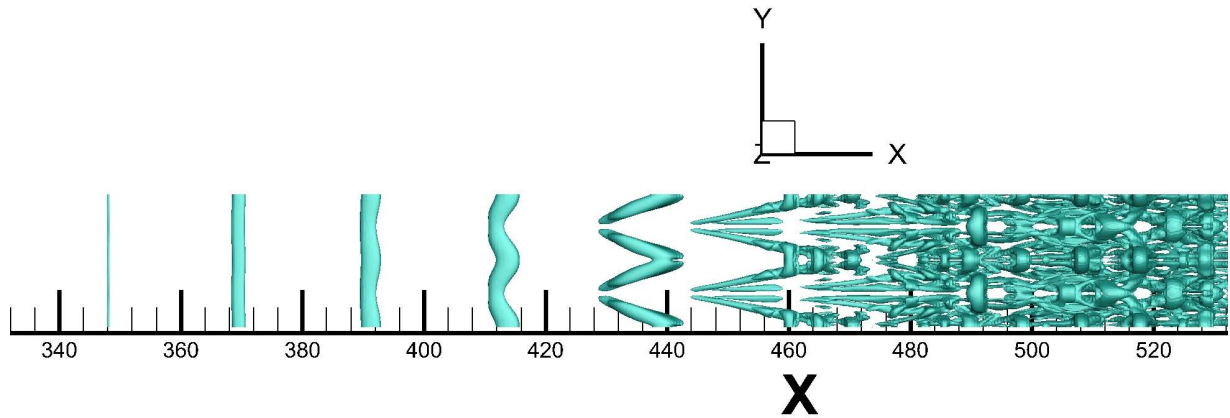

**Figure 7.** The evolution of vortex formation on flow transition by the modified Omega-Liutex method with iso-surface of $\widetilde{\Omega}_L = 0.51$.

In our study, the new concept of vortex core line is applied as the more precise and reasonable way to demonstrate the formation of hairpin vortex without use of iso-surfaces. Therefore, no specified threshold is required. As shown in figure 8 (Left), the evolution of the Λ-vortex to the hairpin vortex is captured by Liutex magnitude of $R = 0.05$. Then, with the convenient tool of Liutex method, Liutex magnitude gradient lines are plotted as shown in figure 8 (Right). To uniquely represent the vortex, the vortex core lines are visualized and colored by Liutex magnitudes as shown in figure 9. The Λ-vortex is shown with iso-surface at $t = 6.0T$ and starts to form the first ring in figure 9 (Left). It is clearly shown by the vortex core lines that the rings, which are on top and in the upstream of the tips, are separated from the legs of the Λ-vortex as shown in figure 9 (Right). Similarly, at $t = 6.3T$ and $t = 6.5T$, the second and third rings are generated separately above the leg of Λ-vortex by the visualization of vortex core lines shown in figure 9 (Right). The Λ-vortex is likely connected with ring-like vortex at the upstream of the tip part when it is observed with iso-surface. However, the unique vortex core demonstrates that the Λ-vortex is apparently separated from the ring-like vortex.

Therefore, our results are consistent with the results of studies [1, 2] in 2014 by DNS observation. It can be concluded as follows:

1) The Λ-vortex and the ring-like vortex are formed separately.
2) The ring-like vortex is not part of Λ-vortex.
3) There is no process that the Λ-vortex self-deforms to the hairpin vortex.

Since the Λ-vortex is not self-deformed to the hairpin vortex, there must be another mechanism of the formation of hairpin vortex. In the next part, it will be explained that the formation of the Λ-vortex to the hairpin vortex is caused by K-H instability.

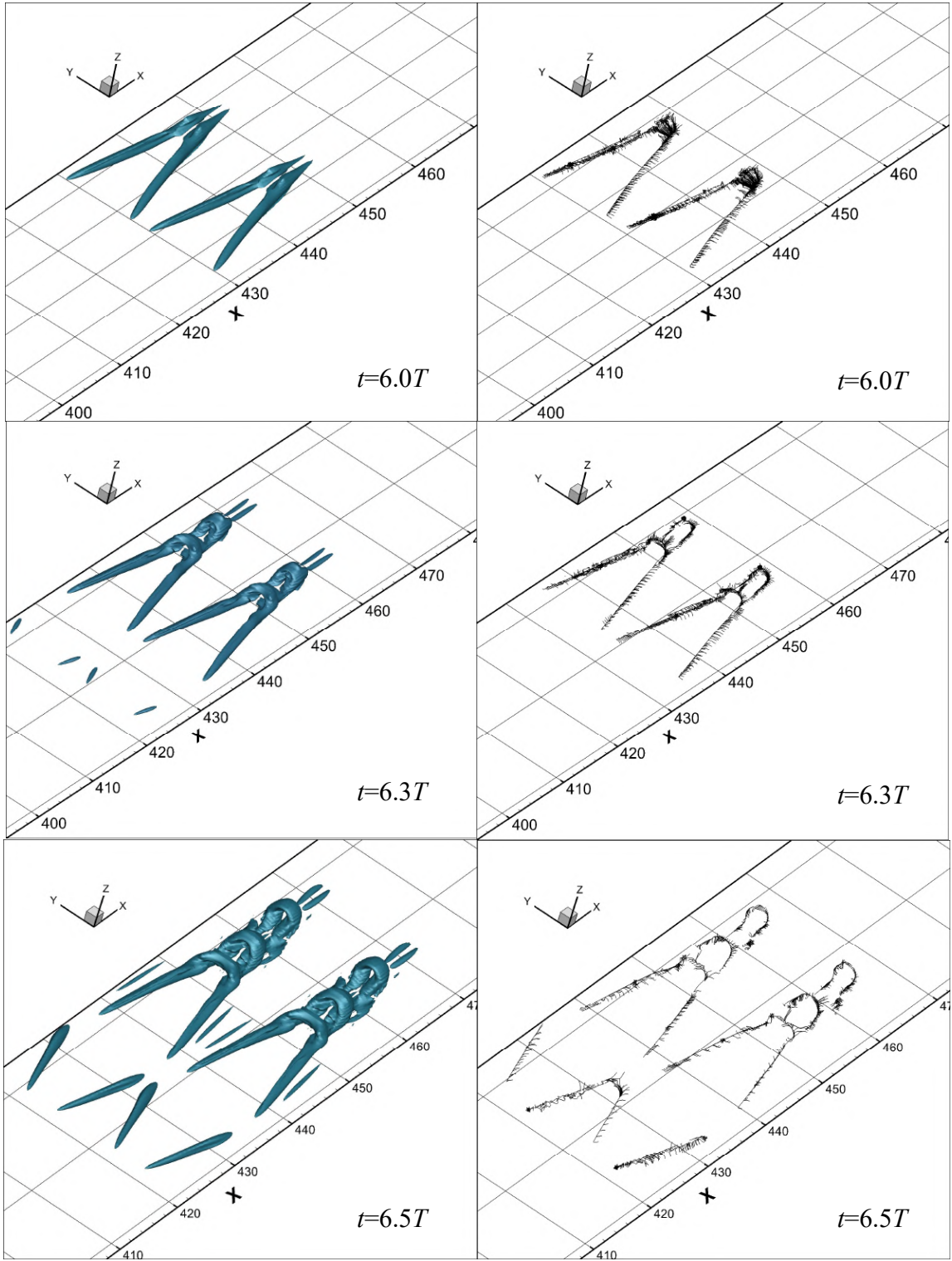

**Figure 8.** The evolution of Λ-vortex and ring-like vortex at $t = 6.0T$, $t = 6.3T$ and $t = 6.5T$.
(Left) Iso-surface of Liutex magnitude $R = 0.05$ (Right) Liutex magnitude gradient lines without iso-surface

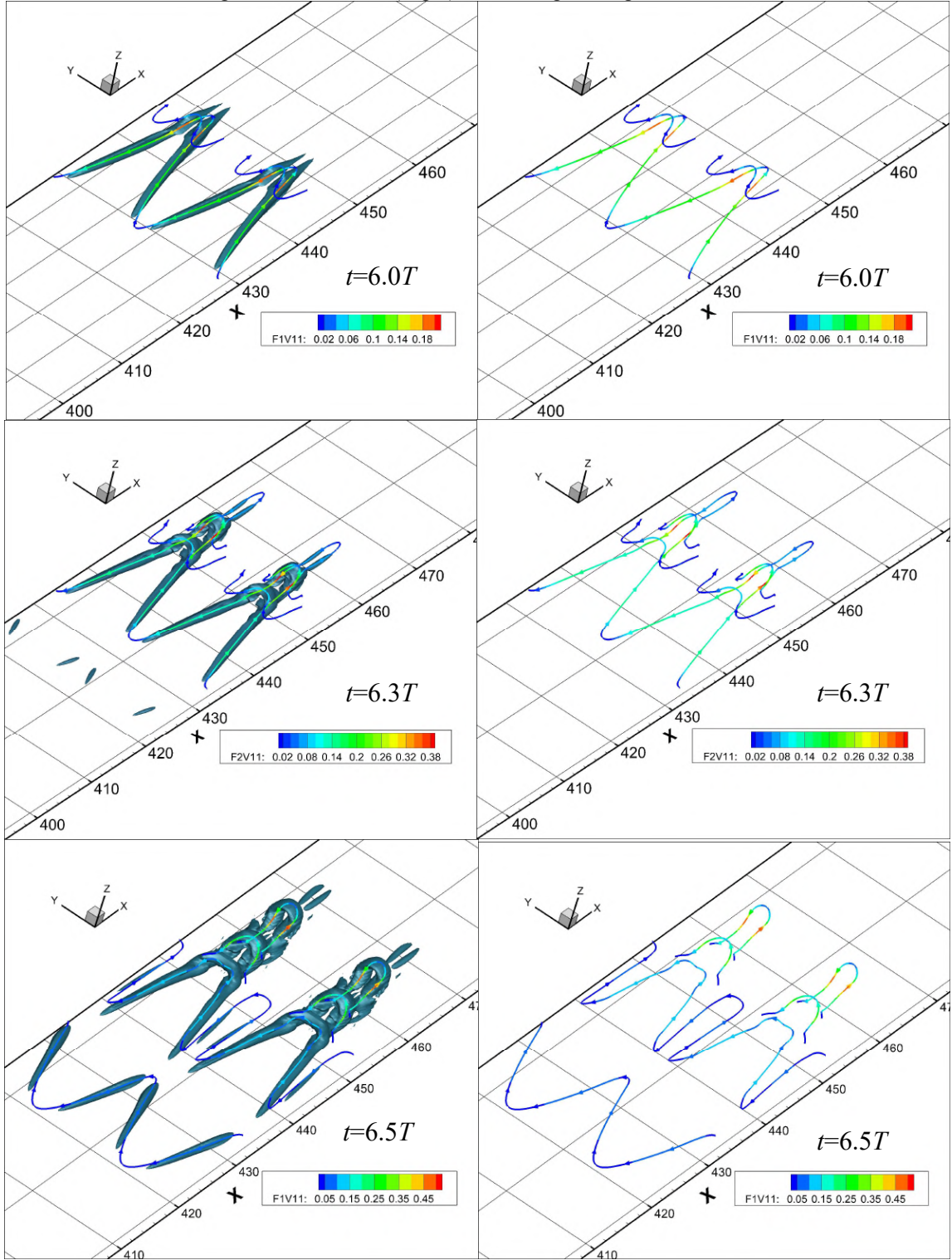

**Figure 9.** The evolution of Λ-vortex and ring-like vortex at $t = 6.0T$, $t = 6.3T$ and $t = 6.5T$
(Left) Iso-surface of Liutex magnitude $R = 0.05$ with the vortex core lines (Right) the vortex core lines without iso-surface colored by Liutex magnitudes

According to the studies [1, 2], the mechanism of ring-like vortex formation has been studied and resulted that the ring-like vortex is formed by K-H instability. It can be described as follows:

1) A momentum deficit zone (low speed zone) is formed above the Λ-vortex and further generates a Λ-shaped high shear due to the Λ-vortex root ejection. The first vortex ring is generated by the high shear layer (K-H type) instability near the tip of the Λ-structure.
2) Multiple vortex rings are all formed by shear layer instability which is generated by momentum deficit.
3) If there is a shear layer inside the flow field, the shear must transfer to rotation and further to turbulence when the Reynolds number is large enough.
4) All small vortices are generated by shear layer.
5) The multiple level shear layers are generated by vortex sweeps and ejections. The sweep brings high speed flow down (positive spike) to the lower boundary layer and the ejection brings the low speed flow up (negative spike) to the upper boundary layer. They form the multiple level shear layers.
6) The shear layer instability (K-H instability) plays an important role to form the ring-like vortex.

From the above analysis, we can conclude that the self-deformation theory of hairpin vortex formation is produced by misunderstanding that considers "vortex" as a "vorticity tube". However, vortex and vorticity are two different concepts and one cannot use vorticity to measure vortex.

**Kelvin–Helmholtz instability**

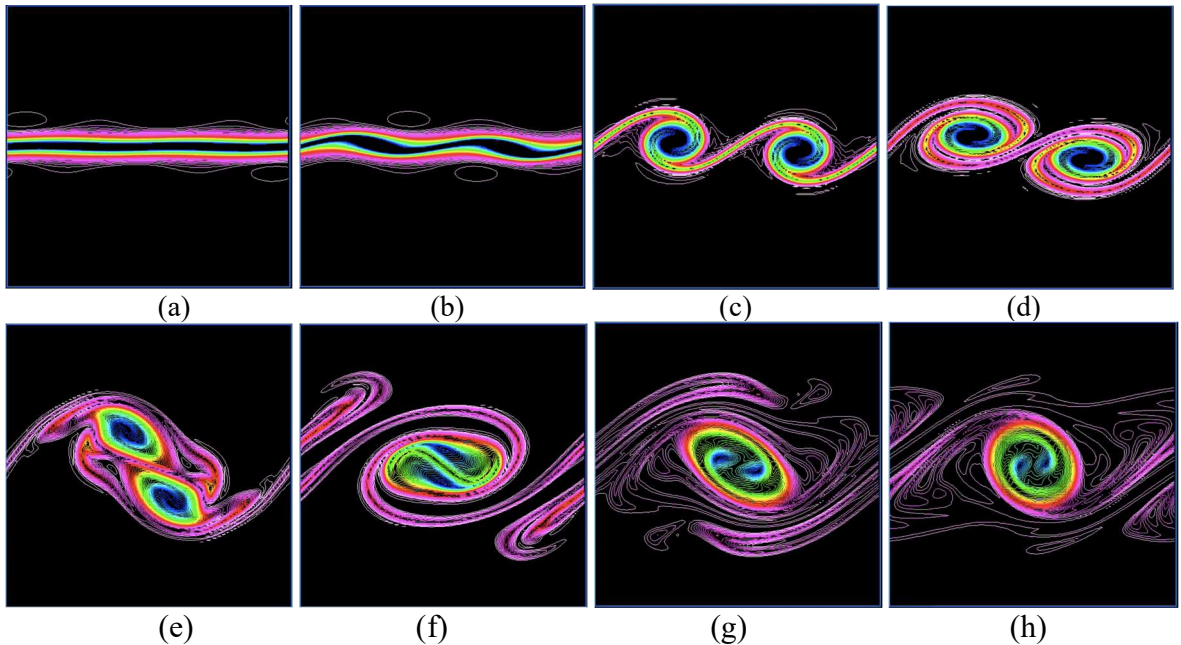

**Figure 10.** Numerical simulation in 2-D on K-H instability (Start from (a) to (h), respectively).

The Kelvin-Helmholtz instability occurs when there is a shear velocity at the interface between two fluid layers. A 2-D vortex ring formation, which is caused by the K-H instability, can be explained as shown in figure 10. A low speed (on the bottom) and a high speed region (on the top) in opposite directions generate shear layer as shown in figures 10(a)-10(b). Then, the shear layer instability (K-H instability) forms the ring-like vortex by transferring the shear to a pair of rotations as in figures 10(c)-10(d). In other words, it is a process to transfer non-rotational vorticity or shear to rotational vorticity or Liutex. After that the pair of rotations becomes one but still have a pair of cores inside the rotation as shown in figures 10(e)-10(h). Therefore, pairing of vortices is an outstanding symbol of K-H instability. Keep in mind that although K-H instability is for the inviscid flow, it could be used to describe about "the shear layer instability" for the viscous flow in our study.

In the case of pairing in K-H instability, we propose a hypothesis that the vortical structures and some features in the same pairs should be similar. Hence, the POD is applied to show results that support the hypothesis and analyze the ring-like vortex formation. In the next section, the review of POD method will be first described, and then the results of POD analysis will be presented.

## 5. POD analysis on K-H instability in flow transition

The POD is applied to extract the coherent structure of complex flows and investigate the evidence of the flow transformation. The $n$ snapshots $x_i = p(\xi, t)$ are assembled columnwise in a matrix $P \in \mathbb{R}^{m \times n}$, which is set from our DNS data by

$$P = \begin{bmatrix} | & | & & | \\ x_1 & x_2 & \cdots & x_n \\ | & | & & | \end{bmatrix}_{m \times n} \tag{1}$$

over three-dimensional discrete spatial points $\xi$ at times $t$. In other words, $m$ is the number of spatial points and $n$ is the number of snapshots (time steps). In our case, $n \ll m$.

In POD method, the flow field is decomposed into a set of basis functions and mode coefficients as

$$p(\xi, t) = \sum_{i=1}^{n} \phi_i(\xi) a_i(t)$$

or it can be written in the matrix form as

$$P = \Phi A \tag{2}$$

where $\Phi$ contains the spatial modes $\phi_i(\xi)$ and $A$ contains the temporal amplitudes $a_i(t)$.

Then, the covariance matrix $C$ is constructed by $C = P^T P$, and the eigenvectors $\psi_i$ and eigenvalues $\lambda_i$ of $C$ can be obtained by

$$C\psi_i = \lambda_i \psi_i, \quad i = 1, 2, \ldots, n.$$

Since $C$ is symmetric and positive-semidefinite, $\lambda_i$ are real and nonnegative with $\lambda_1 \geq \lambda_2 \geq \cdots \geq \lambda_n \geq 0$. Also, the eigenvectors $\boldsymbol{\psi}_i$ are orthogonal. Then, the POD spatial mode $\boldsymbol{\phi}_i$ can be recovered by

$$\boldsymbol{\phi}_i = P\boldsymbol{\psi}_i \frac{1}{\sqrt{\lambda_i}}, \quad i = 1,2,\ldots,n$$

or it can be written in the matrix form as

$$\Phi = P\Psi W, \tag{3}$$

where $W$ is an $m \times n$ matrix with all elements zero except for those along the diagonal. The diagonal elements of $W$ consist of $\frac{1}{\sqrt{\lambda_i}}$.

As our DNS data is a discrete set, the POD is applied as a matrix decomposition called the singular value decomposition (SVD). The studies [39, 40] proposed that POD algorithm is processed in the same way as SVD. By Eq.(3), the matrix $P$ can be decomposed to

$$P = \Phi S \Psi^T. \tag{4}$$

Here, $T$ stands for the transpose of the matrix, $\Phi$ is an $m \times m$ orthogonal matrix, and $\Psi$ is an $n \times n$ orthogonal matrix. The matrix $S$ is an $m \times n$ matrix with all elements of zero except for those along the diagonal. The diagonal elements of $S$ consist of $S_{ii} = \sigma_i \geq 0$ and $\sigma_i$, which represents the kinetic energy of fluid flow, is called a singular values of $P$ with $\sigma_1 \geq \sigma_2 \geq \cdots \geq \sigma_n$. The singular values of $P$ relate to the eigenvalues of matrix $P^t P$ by $\sigma_i = \sqrt{\lambda_i}$. The matrices $\Phi, \Psi$ and $S$ are demonstrated as follows.

$$\Phi = \begin{bmatrix} | & | & & | \\ \boldsymbol{\phi}_1 & \boldsymbol{\phi}_2 & \cdots & \boldsymbol{\phi}_m \\ | & | & & | \end{bmatrix}, \quad S = \begin{bmatrix} \sigma_1 & 0 & \cdots & 0 \\ 0 & \sigma_2 & \cdots & 0 \\ \vdots & \vdots & \ddots & \vdots \\ 0 & 0 & 0 & \sigma_n \\ 0 & 0 & 0 & 0 \\ \vdots & \vdots & \vdots & \vdots \\ 0 & 0 & 0 & 0 \end{bmatrix}, \quad \Psi = \begin{bmatrix} | & | & & | \\ \boldsymbol{\psi}_1 & \boldsymbol{\psi}_2 & \cdots & \boldsymbol{\psi}_n \\ | & | & & | \end{bmatrix}.$$

By the dimension reduction, for $r \leq n$ the reconstructed matrix $P$ can be expressed as

$$P \approx \begin{bmatrix} | & | & & | \\ \boldsymbol{\phi}_1 & \boldsymbol{\phi}_2 & \cdots & \boldsymbol{\phi}_r \\ | & | & & | \end{bmatrix}_{m \times r} \begin{bmatrix} \sigma_1 & 0 & \cdots & 0 \\ 0 & \sigma_2 & \cdots & 0 \\ \vdots & \vdots & \ddots & \vdots \\ 0 & 0 & 0 & \sigma_r \end{bmatrix}_{r \times r} \begin{bmatrix} - & \boldsymbol{\psi}_1 & - \\ - & \boldsymbol{\psi}_2 & - \\ & \vdots & \\ - & \boldsymbol{\psi}_r & - \end{bmatrix}_{r \times n}. \tag{5}$$

Consequently, the reconstructed matrix $P$ in Eq.(5) can also be written into a linear combination as

$$P \approx \sum_{i=1}^{r} \sigma_i \boldsymbol{\phi}_i \boldsymbol{\psi}_i^T \tag{6}$$

To choose the size $r$ of the reduced dimension of matrix, it can be evaluated from the relative energy of the snapshots by the first $r$ POD basis vectors using the following formula:

$$\epsilon(r) = \frac{\sum_{i=1}^{r} \lambda_i}{\sum_{i=1}^{n} \lambda_i}, \tag{7}$$

where $1 - \epsilon(r) \leq tol$ with $0 < tol < 1$.

After the dimension reduction is performed, the whole structure can be extracted into $r$ coherent structures. The basis $\boldsymbol{\phi}_i$ in Eq.(6) is a POD spatial mode. Dimensional reduction can keep the most important mode as a basis. The first mode will be the most dominant energy structure and the last mode will be the least dominant energy structure.

By our DNS data, the vortex structure of flow on a flat plate is investigated. Figure 11 shows one example of vortex in flow transition at $t = 13.00T$, where $T$ is a period of T-S waves along the streamwise direction between $x = 430$ and $x = 630$, visualized by using the modified Omega-Liutex method with iso-surface of $\widetilde{\Omega}_L = 0.52$.

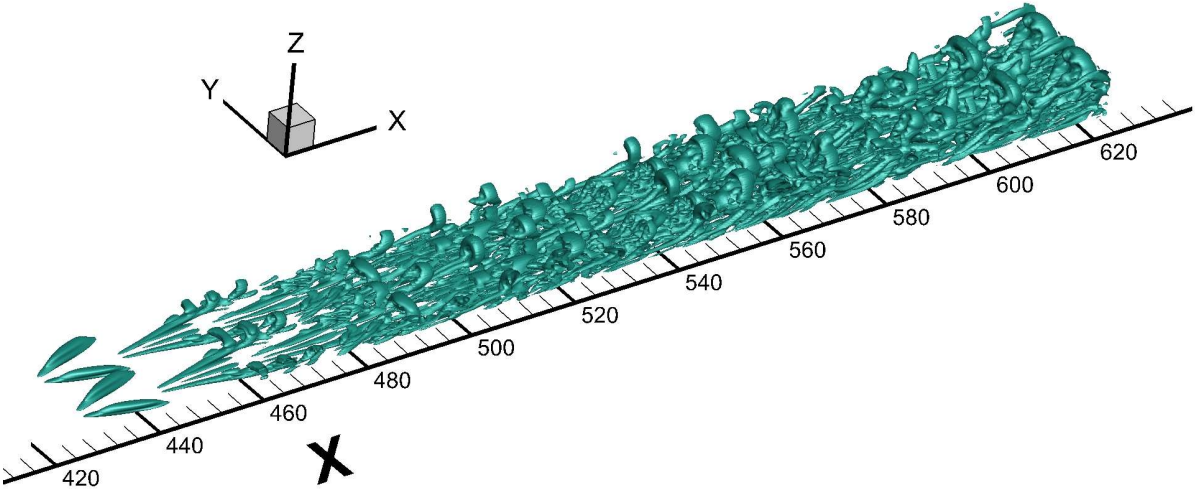

**Figure 11.** Vortex structures by modified Omega-Liutex method with $\widetilde{\Omega}_L = 0.52$ at $t = 13.00T$

In order to apply POD in the analysis of K-H instability, the domain of study is the area where a $\Lambda$-vortex becomes a hairpin vortex. Therefore, the specific area of POD analysis is from $x = 460$ to $x = 490$ as shown in figure 11. The POD is applied over 100 snapshots (i.e. 100 time steps) at time between $t = 12.51T$ and $t = 13.50T$ to investigate the principal components of the coherent structures. A subzone is extracted to reduce the computation complexity. The parameters of the subzone are given in Table 2.

|  | Start Index | End Index |
|---|---|---|
| I (in $x$ direction) | 316 | 376 |
| J (in $y$ direction) | 1 | 128 |
| K (in $z$ direction) | 1 | 200 |

**Table 2.** Parameters of subzone

The snapshot $x_i$ is defined to install into the matrix $P$ in Eq. (1) by

$$x_i = \begin{pmatrix} u^{(i)}_{316,1,1} \\ \vdots \\ u^{(i)}_{376,1,1} \\ u^{(i)}_{316,2,1} \\ \vdots \\ u^{(i)}_{376,2,1} \\ \vdots \\ u^{(i)}_{I,J,K} \\ \vdots \\ u^{(i)}_{316,128,200} \\ \vdots \\ v^{(i)}_{I,J,K} \\ \vdots \\ w^{(i)}_{I,J,K} \\ \vdots \\ w^{(i)}_{376,128,200} \end{pmatrix} \quad \text{for } i = 1,\dots,100,$$

where $u^{(i)}, v^{(i)}$ and $w^{(i)}$ are velocity fields at $t = (12.50 + 0.01i)T$, for $i = 1,\dots,100$.

    The POD is applied over 100 snapshots in our study. Therefore, 100 POD modes are obtained. The eigenvalues in POD method are evaluated and ordered. The POD modes are ranked by the energy content represented by the singular values or the eigenvalues. The results of eigenvalues of all modes are investigated. It is found that two eigenvalues are similar to each other in pairs. The eigenvalues of mode 1 and the pairs of some other modes are presented as shown in Table 3.

| Pair No. | Eigenvalues of modes | |
|---|---|---|
|  | Mode 1: 87949612.7013267 | |
| Pair 1 | Mode 2: 200288.009080135 | Mode 3: 194889.125828145 |
| Pair 2 | Mode 4: 39234.2518602893 | Mode 5: 38676.5370753112 |
| Pair 3 | Mode 6: 34451.0788998119 | Mode 7: 34089.7551347020 |
| Pair 4 | Mode 8: 18673.8222690667 | Mode 9: 18322.4804874360 |
| **Pair 5** | Mode 10: 17169.6248912740 | Mode 11: 17018.6436759498 |
| **Pair 6** | Mode 12: 11631.8322897197 | Mode 13: 11494.4484482407 |
| **Pair 7** | Mode 14: 8146.26652550679 | Mode 15: 8090.41774555481 |
| **Pair 8** | Mode 16: 2844.71672727352 | Mode 17: 2795.57150355193 |
| **Pair 9** | Mode 18: 1452.34341004029 | Mode 19: 1397.08863389295 |
| **Pair 10** | Mode 20: 1201.55853843886 | Mode 21: 1196.85608390518 |
| **Pair 11** | Mode 34: 6.10591736207254 | Mode 35: 5.27626505327566 |

**Table 3.** Eigenvalues of some modes

In figure 12, mode 1, which represents the mean velocity (streamwise velocity), is taken out and only the fluctuation modes, which are mode 2 to mode 100, are considered. The eigenvalue of each mode excluding mode 1 and the pairings of eigenvalues can be seen in figure 12.

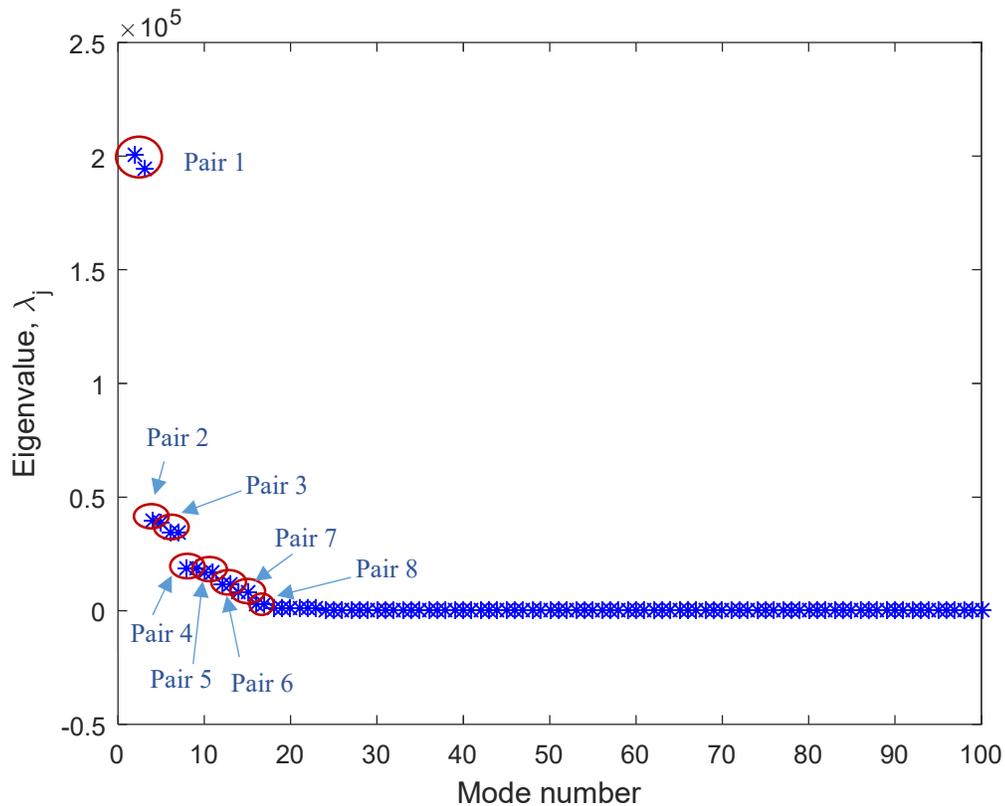

**Figure 12.** Eigenvalues of mode 2 to mode 100 indicated the pairs of modes.

By dimensional reduction, the number of modes, $r$, is chosen by Eq. (7) and shown in figure 13. The proper number of modes to reconstruct the vortex structure is 20 modes since it can keep the most cumulative energy at about 100%.

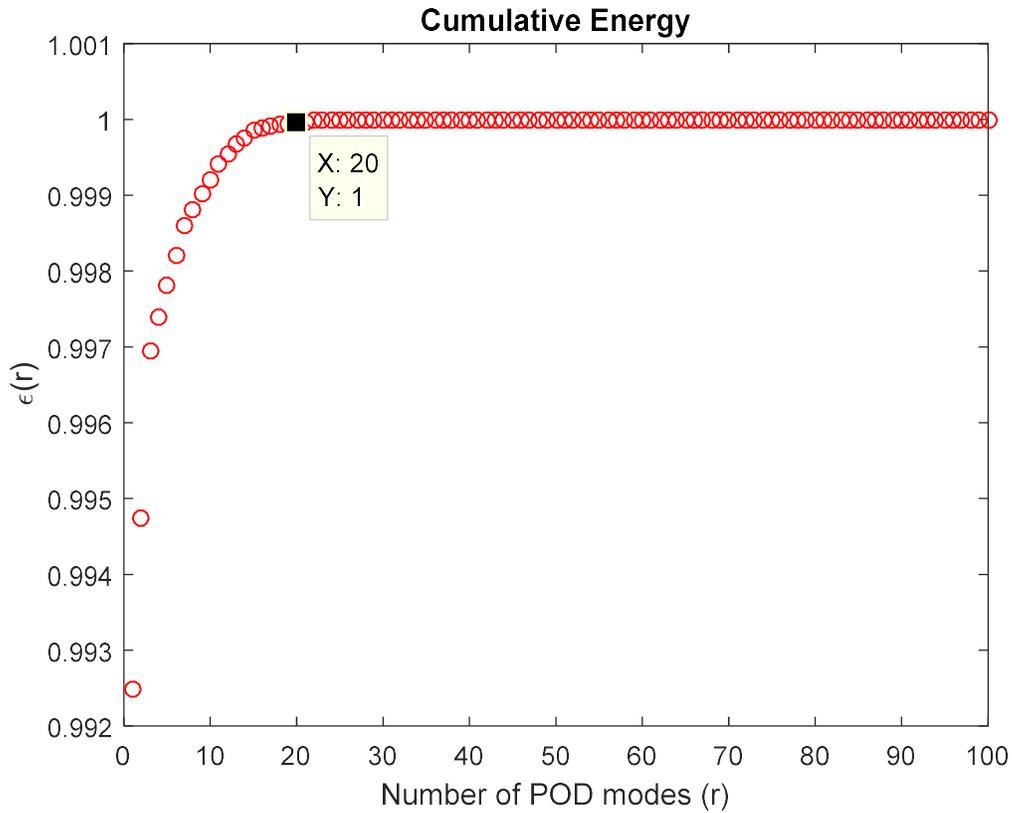

**Figure 13.** Distribution of $\epsilon(r)$

The reconstruction performs very well with the first 20 modes comparing to the original flow data. Figure 14 demonstrates that the vortex structures of the original flow and the reconstructed flow of first 20 modes in three different time steps.

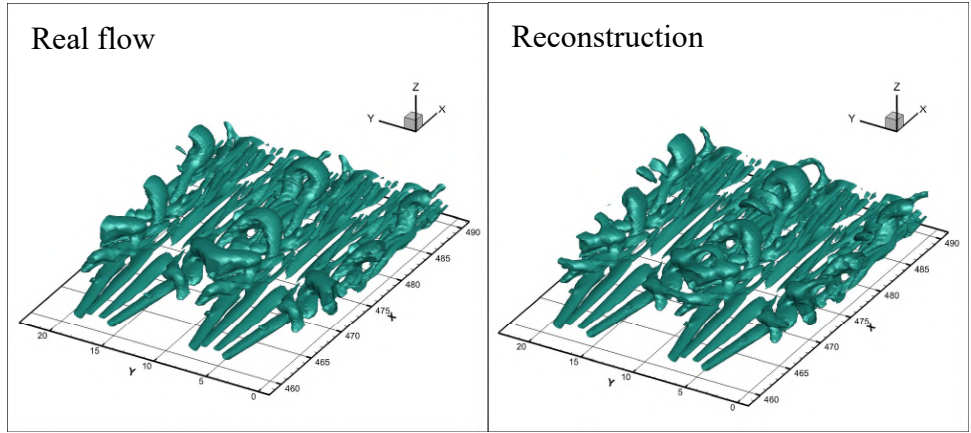

(a) The vortex structure at $t = 12.51T$

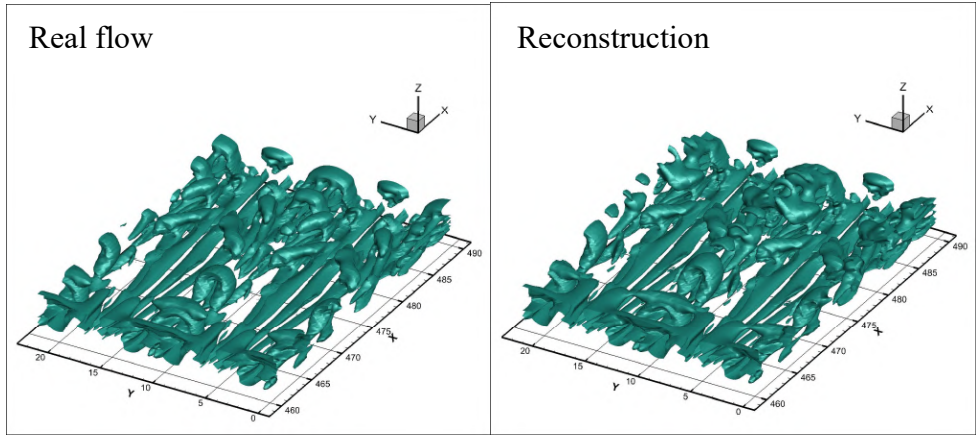

(b) The vortex structure at $t = 13.00T$

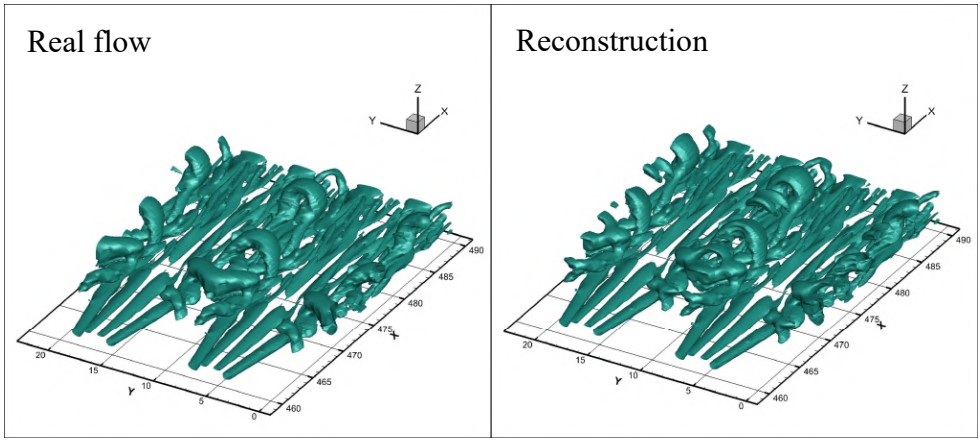

(c) The vortex structure at $t = 13.50T$

**Figure 14.** The vortex structures of real flows (left) and reconstructions of the first 20 modes (right) with iso-surface of $\widetilde{\Omega}_L = 0.52$ with $\varepsilon = 0.001(\beta^2 - \alpha^2)_{max}$.

The vortex structures in both original DNS results and reconstructed ones are in different shapes in different time steps as shown in figure 14. However, the same shape of vortex structures of each POD mode in different time step from $t = 12.51T$ to $t = 13.50T$ is obtained, which is really the eigenvector, since the POD mode represents a spatial structure of the flow with the time average.

Mode 1 represents the most dominant mode in terms of kinetic energy content as it has the most principal component with $\epsilon(r) = 99.25\%$. In figure 15, mode 1 is shaped to streamwise vortex structure with the iso-surface of $\widetilde{\Omega}_L = 0.52$. The spatial shapes of the other modes are also illustrated in pairs which can be seen in figure 15. The shapes of modes in the same pair are similar as they have similar eigenvalues and similar eigenvectors.

For modes in pair 1, which is a pair of modes 2 and 3, are likely in streamwise characteristic visualized by modified Omega-Liutex method with iso-surface of $\widetilde{\Omega}_L = 0.52$. For the other modes, the vortex shapes are dominated by the spanwise structures. As we can see in figure 15, the higher modes have smaller spanwise vortex structures and more numbers of rings visualized by the modified Omega-Liutex method.

| Pair No. | POD Mode Shape | |
|---|---|---|
| | Mode 1 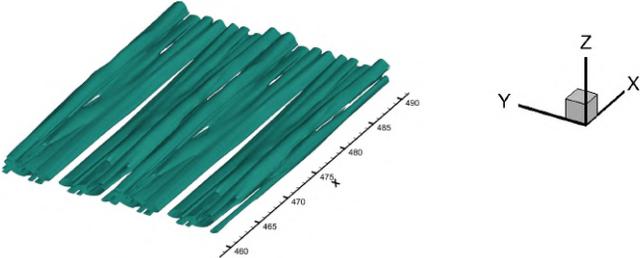 | |
| Pair 1 | Mode 2 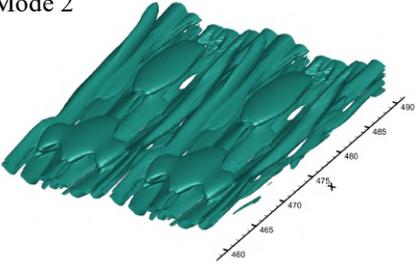 | Mode 3 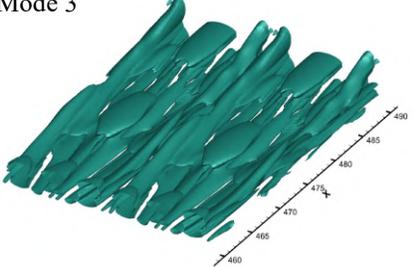 |
| Pair 2 | Mode 4 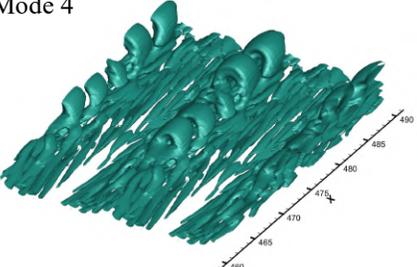 | Mode 5 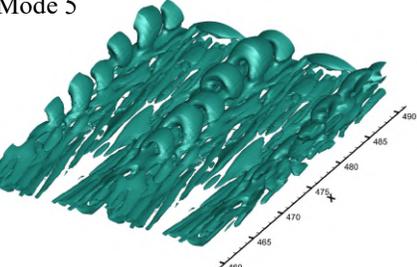 |
| **Pair 3** | Mode 6 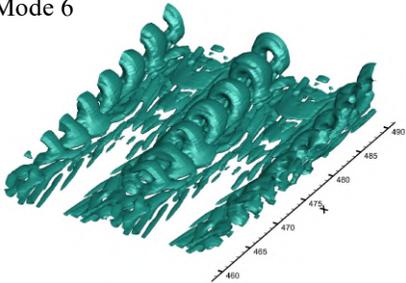 | Mode 7 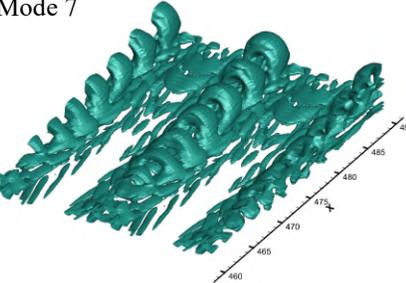 |
| **Pair 4** | Mode 8 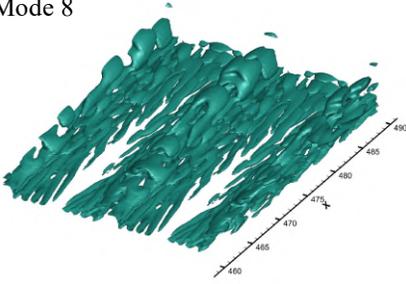 | Mode 9 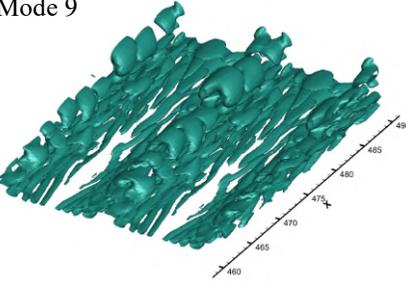 |
| **Pair 5** | Mode 10 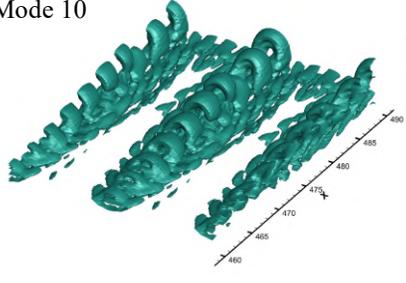 | Mode 11 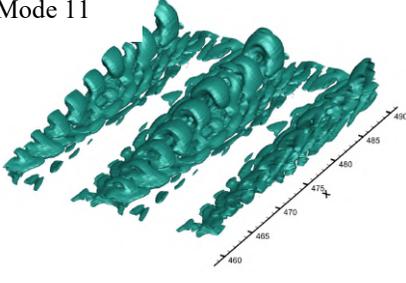 |
| **Pair 6** | Mode 12 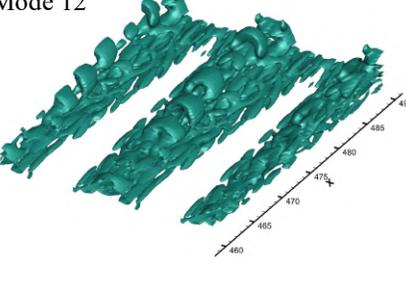 | Mode 13 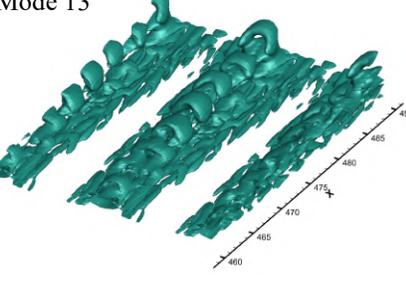 |
| **Pair 7** | Mode 14 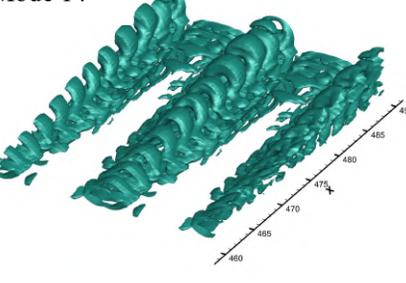 | Mode 15 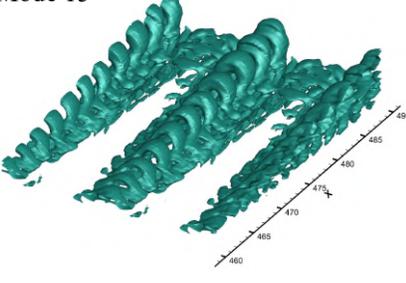 |

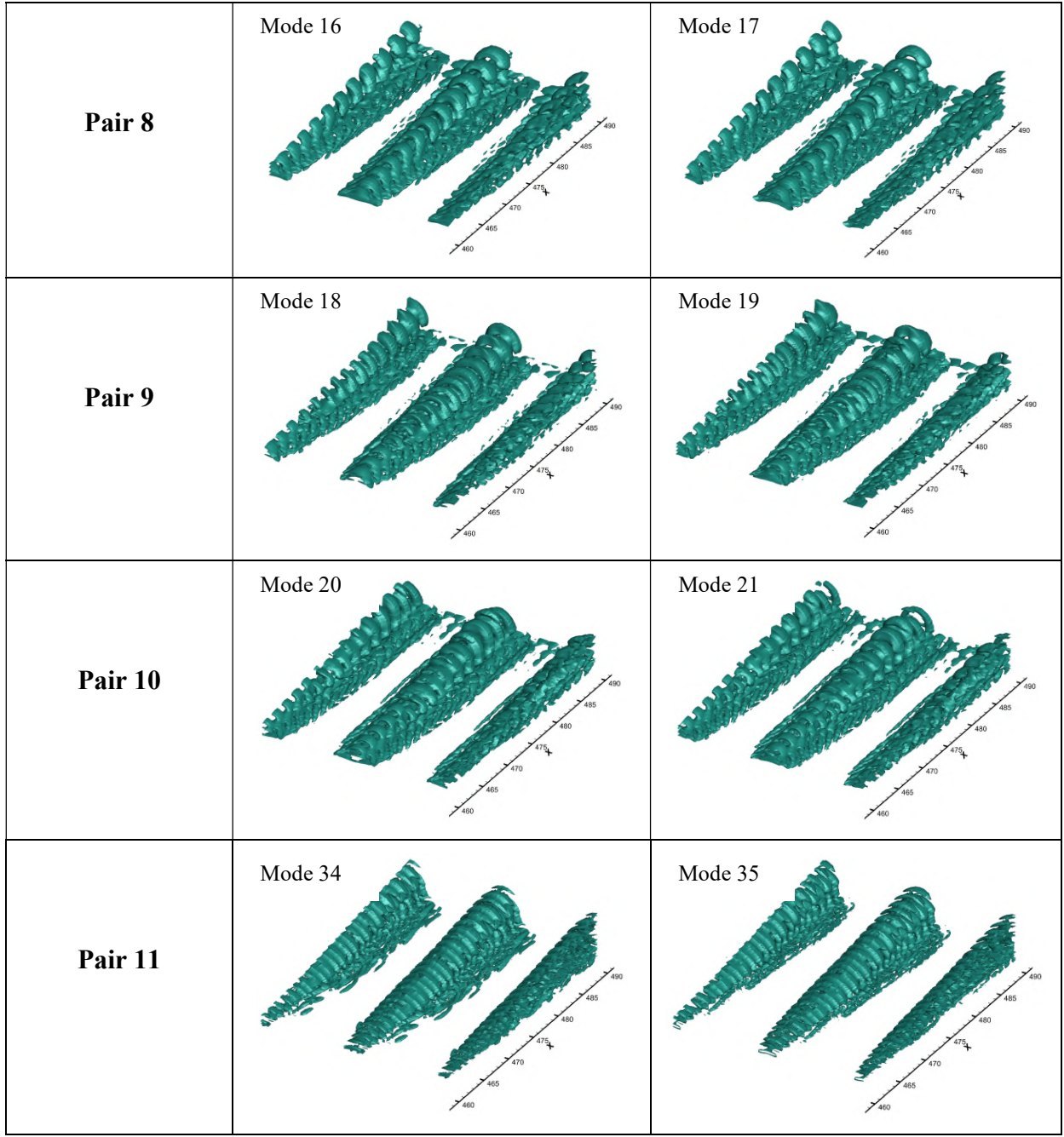

**Figure 15.** Vortex structures of some modes with iso-surface of $\widetilde{\Omega}_L = 0.52$ with $\varepsilon = 0.001(\beta^2 - \alpha^2)_{max}$.

To investigate the fluctuation of flows, the POD time coefficients are applied to demonstrate periodicities of POD modes. The POD time coefficients are scaled by the singular values of the eigenvectors $\boldsymbol{\psi}_i$ and they can be obtained from the matrix of temporal structure $A = S\Psi^T$, which is obtained from Eqs.(2) and (4). As we can see in figure 16, there is no fluctuation in mode 1 since the graph of POD time coefficient is likely constant. Thus, mode 1 can represent the mean flow. For the other modes, time coefficients are illustrated in pairs. The similar fluctuations features of modes as periodicities and amplitudes are demonstrated in the same pairs. We can see that the higher modes have higher fluctuations as shown in figure 16.

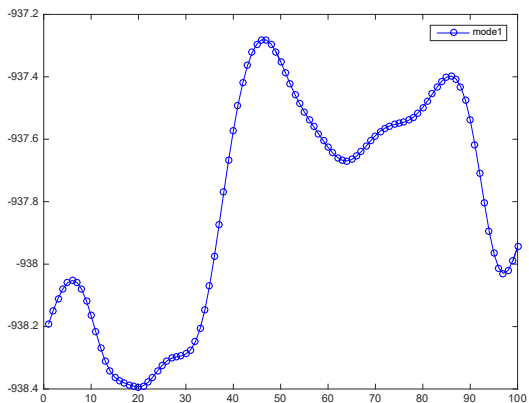

Mode 1

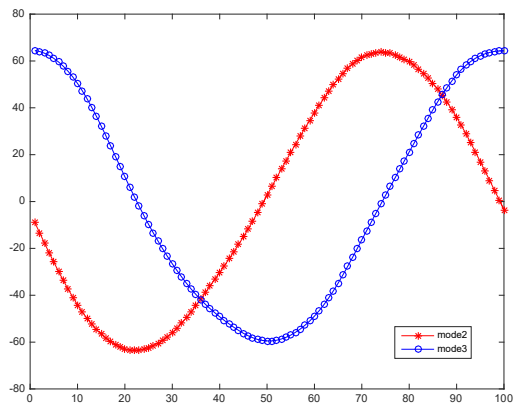

Mode 2 (Red) and Mode 3 (Blue)

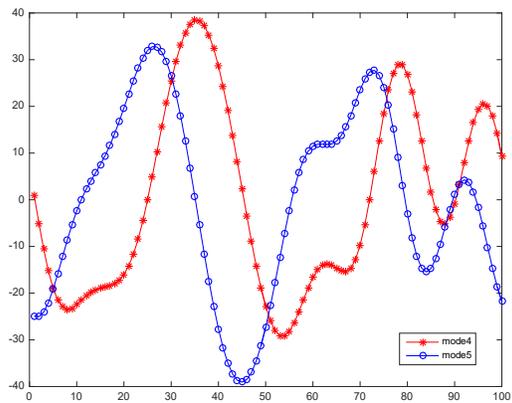

Mode 4 (Red) and Mode 5 (Blue)

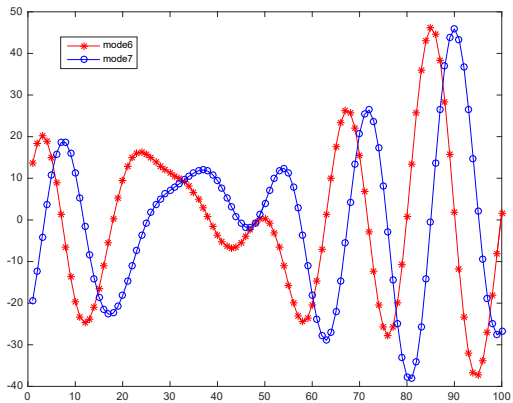

Mode 6 (Red) and Mode 7 (Blue)

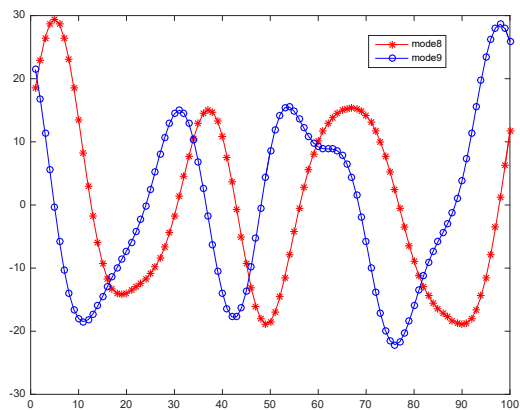

Mode 8 (Red) and Mode 9 (Blue)

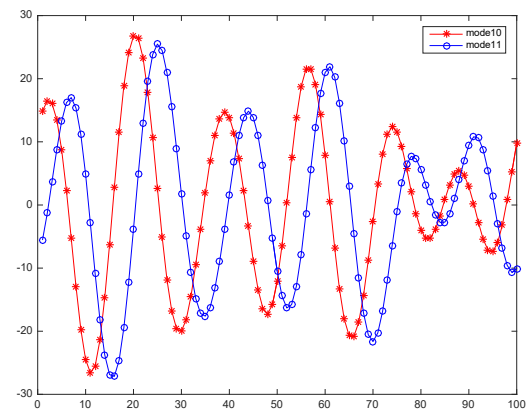

Mode 10 (Red) and Mode 11 (Blue)

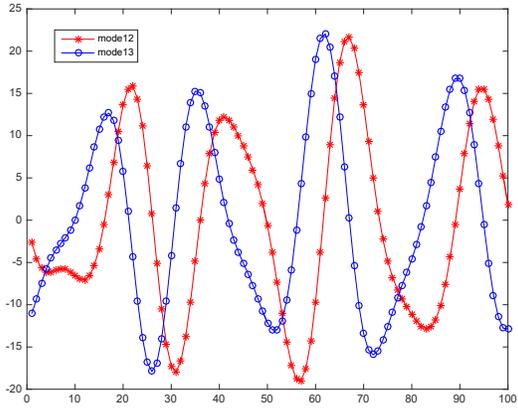
Mode 12 (Red) and Mode 13 (Blue)

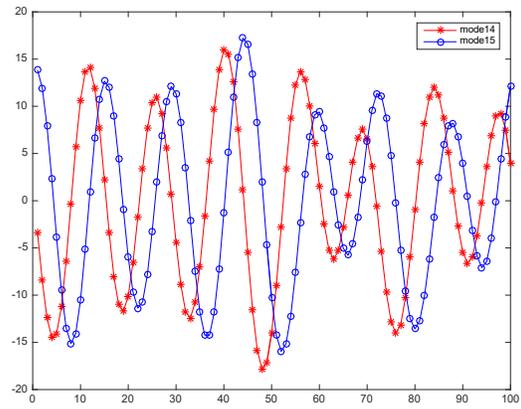
Mode 14 (Red) and Mode 15 (Blue)

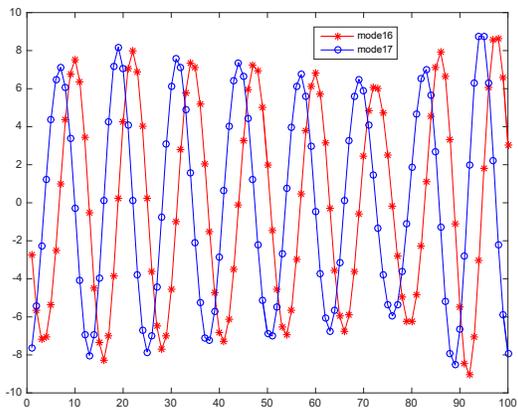
Mode 16 (Red) and Mode 17 (Blue)

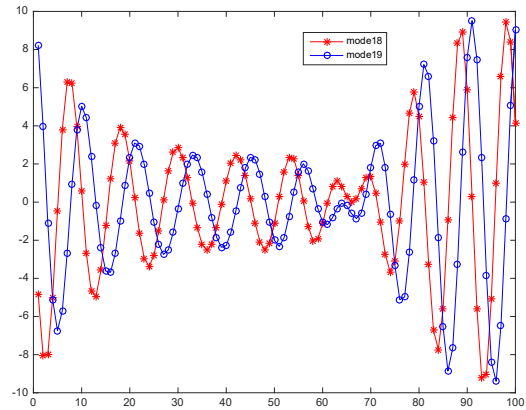
Mode 18 (Red) and Mode 19 (Blue)

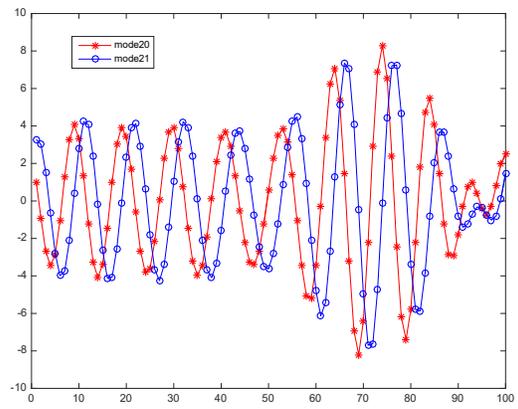
Mode 20 (Red) and Mode 21 (Blue)

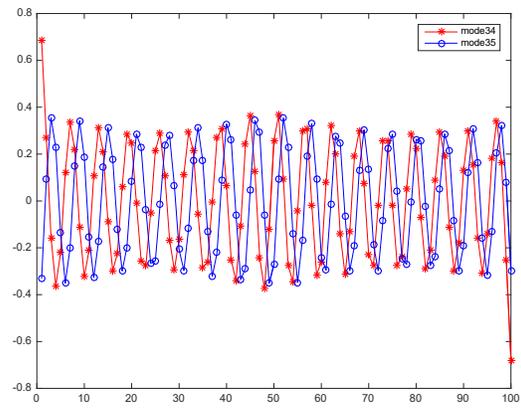
Mode 34 (Red) and Mode 35 (Blue)

**Figure 16.** POD time coefficients (y-axis) with mode number (x-axis)

## 6. Conclusions

The conclusions of this paper are obtained into two main results. First, the vortex core line based on Liutex definition, which represents the rotation axis, is applied to show the formation of hairpin vortex. The results show that the ring-like vortex is not part of the Λ-vortex and they are formed separately. Second, the POD is applied to a DNS data with 100 snapshots to analyze the vortex structure and K-H instability characteristics in boundary layer flow transition where the area of a Λ-vortex becomes a hairpin vortex. Mode 1 is the most energetic mode with leading streamwise structure and no fluctuation. The other modes are in spanwise shapes with fluctuations. The higher modes have smaller spanwise vortex structures as they are ranked by the eigenvalues. The smaller eigenvalues in higher modes get the smaller scale structures. The higher modes, which are dominated by the spanwise structures, show more fluctuations. Moreover, the important result shows that the eigenvalues and eigenvectors of POD modes are shown in pairs, which is typical in the K-H instability. The same characteristics such as mode shapes and fluctuations are also established in pairs. This evidence is a strong proof that the K-H instability plays a key role in hairpin vortex formation. K-H instability is always displayed in pairs. Therefore, it can be strongly confirmed that the K-H instability is the main factor of the formation of the hairpin vortex from the Λ-vortex, which transform the non-rotational vorticity or shear to the rotational vorticity or Liutex. The theory of self-deformation of the Λ-vortex to hairpin vortex has contradiction to our DNS observation and analysis and thus may have no scientific foundation.

## Acknowledgments

The authors thank the Department of Mathematics of University of Texas at Arlington and Royal Thai Government for the financial support. The authors are grateful to TACC (Texas Advanced Computation Center) for providing CPU hours to this research project. The computation is performed by using Code DNSUTA which was released by Dr. Chaoqun Liu at University of Texas at Arlington in 2009.